\documentclass[sigconf,nonacm,balance=false]{acmart}
\setcopyright{none}
\hypersetup{pdfauthor={Zihao Sun},pdfcreator={LaTeX},pdfproducer={LaTeX}}
\usepackage{booktabs}
\usepackage{flushend}
\AtBeginEnvironment{thebibliography}{\interlinepenalty=10000}
\begin{document}

\title[From Source-Based Quizzes to Reteaching Decisions]{From Source-Based Quizzes to Reteaching Decisions: A Practitioner Guide to ClassQuest}
\author{Zihao Sun}
\affiliation{\institution{University of Technology Sydney}\city{Sydney}\country{Australia}}
\email{sun.zihao@student.uts.edu.au}
\begin{abstract}
A source-based quiz can help a computing teacher identify material to explain again, but its report must preserve the limits of the evidence behind that decision. We present ClassQuest, a prototype connecting teacher-reviewed question drafts, private student responses, topic reports and targeted retries. A Year 10 classroom account of an earlier prototype supplies 20 usefulness ratings from 24 invitees, with a mean of 8.0 and median of 9 on a ten-point scale. A Boolean-logic episode documents how immediate feedback, a teacher explanation and a fresh question were combined. We separate this account from a completed technical audit and reusable teaching guide. An exhaustive single-item audit found no count mismatches across 5,456 constructed response configurations, while diagnostic cases exposed a retry increase caused entirely by changing participation and a progression suggestion based on one learner's three answers. A further case shows that an unoffered retry can be labelled missing. A separate reporting companion distinguishes opportunity to respond, pending review and matched retry cohorts. Its completion-sensitivity check matched an independent oracle in 46,375 constructed states across roster sizes 1--30, identifying when unknown responses can change a reporting band. Our contributions are a source-linked formative workflow, executable examples of interpretation failures, and reporting and teaching materials for examining those failures. Neither the questionnaire nor the audit establishes improved learning; the findings identify reporting requirements and practical directions for formative teaching.
\end{abstract}
\maketitle

\section{Teaching problem and contribution}
After a short quiz on Boolean logic, a loop, or an AI concept, a teacher must decide what to explain again. The question is not simply whether the class percentage increased. It is which learners supplied evidence, which responses have been judged, what misunderstanding an answer reveals, and whether a later question gives comparable evidence. A report can calculate its percentages correctly while obscuring one of these distinctions.

ClassQuest is designed around this formative teaching cycle. A teacher supplies source material, reviews draft questions and their evidence, opens a private first attempt, examines the class report, discusses a difficulty, and offers a targeted retry. The prototype makes source passages available alongside items and distinguishes missing responses from answers awaiting review. These are useful affordances, but their presence does not ensure that a suggested teaching action is warranted.

The paper contributes three resources for computing educators. \textbf{C1: an adoptable formative-assessment practice} connects source review, individual response, explanation and a fresh question, grounded in a Year 10 Boolean-logic episode and student feedback. \textbf{C2: diagnostic cases for evaluating quiz reports} expose five distinct inference problems: source presence versus answer validity, unoffered items versus non-response, scored answers versus learners, incomplete grading versus coverage, and changing retry participation. \textbf{C3: an implemented reporting companion and decision protocol} translate those cases into explicit quantities, a tested check for completion-sensitive reporting bands, and teacher actions. The contribution is this operational connection between classroom practice, executable counterexamples and report interpretation; it does not depend on a new generation model or a claim of improved learning.

This practitioner paper uses three kinds of evidence: current implementation behaviour, constructed software-audit cases, and an initial classroom account supplied by the practitioner. These have different scopes. The implementation shows what the tool supports; the audit tests selected failure mechanisms; the classroom account describes use and experience. We do not infer a deployment of every current feature from a general report of classroom use.

\section{Prior work and positioning}
\subsection{Feedback becomes formative through its use}
Black and Wiliam~\cite{black} locate formative assessment in how evidence informs subsequent instruction. Nicol and Macfarlane-Dick~\cite{nicol} connect feedback with learners' self-regulation and information useful to teachers. This motivates a practical distinction: generating an explanation, displaying a percentage and choosing a teaching response are separate operations. ClassQuest must support the transition between them rather than assume that a report itself completes the formative cycle.

Feedback content also matters. Shute~\cite{shute} distinguishes verification, explanation, hints and worked examples, while Hattie and Timperley~\cite{hattie} analyse why feedback effects depend on its focus and delivery. Sadler~\cite{sadler} emphasises learners' capacity to judge and improve the quality of their work; that account is particularly relevant to the explanations proposed here, beyond binary correctness alone. These foundations justify examining the rationale for an answer and the learner's next action. They do not validate ClassQuest's implementation thresholds.

\subsection{Computing questions and generated explanations}
Automated programming feedback is established. Keuning et al.~\cite{keuning} distinguish feedback functions, generation techniques, adaptability and evaluation across programming tools. Sarsa et al.~\cite{sarsa} investigated generated exercises and code explanations, including their need for instructor adjustment. Doughty et al.~\cite{doughty} compared generated and human-crafted programming multiple-choice questions at ACE. These precedents rule out question generation alone as the contribution of ClassQuest.

Generated explanations and classroom oversight have also been studied directly. Leinonen et al.~\cite{leinonen} compared student and model-generated explanations using student judgments; Roest et al.~\cite{roest} examined next-step programming hints. Qiao et al.~\cite{qiao} reported instructor moderation of generated discussion responses. Denny et al.~\cite{dennyai} place these uses within broader changes to computing education. Our narrower concern is what evidence remains available when an educator uses generated items and subsequent responses to decide what to reteach.

Process and final-product evidence should remain distinguishable. Edwards et al.~\cite{edwards} examine language models for programming-process data, while Qiu et al.~\cite{qiu} scaffold reflective code evaluation through structured AI assistance. ClassQuest's audit therefore treats an answer, an explanation and a retry as different observations. A located source or keyword match is insufficient to certify the semantic validity of all three.

\subsection{Analytics interpretation and teacher action}
Verbert et al.~\cite{verbert} examine learning dashboards and the relationship between visualisation and understanding. Reviews by Schwendimann et al.~\cite{schwendimann} and Matcha et al.~\cite{matcha} identify limitations in dashboard evaluation and grounding in learning processes. We use these findings to motivate careful evaluation of the present prototype, rather than attribute its particular defects to other systems.

Wise and Vytasek~\cite{wise} distinguish designing analytics technology from designing how people interpret and use it. Holstein et al.~\cite{holstein} demonstrate teacher involvement in the design and classroom evaluation of orchestration support. These are close precedents for preserving teacher judgment. Our additional resource is a set of executable reporting cases and a corresponding decision protocol for source-based computing quizzes. Coverage, cohort matching and human review are established principles; their explicit combination makes the limits of this particular report inspectable and teachable.

\subsection{Peer discussion, practice and evidence of learning}
PeerWise~\cite{peerwise} already supports student question creation and discussion. Smith et al.~\cite{smith} used fresh individual concept questions to distinguish understanding after peer discussion from agreement on a repeated answer; Porter et al.~\cite{porter} investigated the approach in computing. Zingaro and Porter~\cite{peerinstruction} further examined instructor intervention. These studies motivate keeping individual attempts, discussion and teacher explanations distinct in the proposed peer extension.

Repeated practice can serve learning as well as assessment: Karpicke and Roediger~\cite{karpicke} demonstrated a retrieval-practice benefit on delayed vocabulary recall. That finding motivates a future retention measure, not a claim that this quiz reproduces the effect. Likewise, Deslauriers et al.~\cite{deslauriers} distinguished measured learning from feeling of learning in undergraduate physics. The ClassQuest questionnaire measures perceived usefulness, a different construct again; it supplies adoption feedback rather than a substitute for learning assessment.

\section{The prototype and an adoptable workflow}
\subsection{Preparation and teacher review}
ClassQuest accepts teaching documents and extracts passages with page or section locations. Draft questions can be proposed through an optional local language model. When generation is unavailable or fails validation, the application labels simpler extractive starters for teacher revision. Every newly generated question requires teacher approval before publication. The software does not establish pedagogical suitability simply because it can locate a quoted passage.

The teacher should review the learning objective, answer key, distractor rationale and source together. A question about Boolean conjunction, for example, should distinguish a logical rule from a student's familiarity with the word ``and.'' A source sentence can be authentic while the answer key is wrong. This is an instructional reason for the approval step, not merely a requirement that a button be pressed.

The current implementation includes multiple-choice, select-all, ordering and short-explanation questions. Closed responses can be scored mechanically against a reviewed key. An open response receives a concept-overlap hint and remains pending until human review. A high overlap score must not be described as semantic correctness: the same words can occur in an incorrect explanation or a negation of the expected rule.

\subsection{First attempt, discussion and retry}
Learners join a classroom session and submit a private first attempt. The teacher can examine per-question and topic-level evidence, including common wrong answers and confidence information. A discussion stage then permits the teacher to revisit a concept before opening a targeted retry. The current reports distinguish missing responses, wrong responses and work awaiting review.

For another educator, the practical value of this sequence is that each phase can answer a different question. The first attempt indicates what a learner can currently do without the subsequent discussion. The discussion exposes or responds to reasoning. A retry provides another observation after that intervention, but its interpretation depends on who participates and how the items relate. A higher retry percentage is not automatically a measure of learning.

Table~\ref{tab:workflow} describes an adoption guide grounded in the current workflow. Its teaching actions are recommendations, separate from the classroom experience reported below. The current implementation and later audit are not used to infer that every feature was deployed during those sessions.

\begin{table*}[t]
\caption{A reusable ClassQuest teaching workflow. Recommended decisions, rather than a claimed record of the earlier classroom deployment.}
\label{tab:workflow}
\begin{tabular}{@{}p{0.16\textwidth}p{0.37\textwidth}p{0.41\textwidth}@{}}
\toprule
Stage & Evidence to inspect & Teacher decision\\
\midrule
Prepare & Learning objective, source passage, proposed key and distractors & Approve, revise or discard the item; check that the source supports the intended interpretation.\\
First attempt & Individual response and confidence; who was offered each item and who responded & Identify a plausible difficulty without treating non-response as an incorrect answer.\\
Review & Pending explanations and common wrong choices & Resolve ungraded work before making a strong topic-level inference.\\
Discuss & Competing explanations and their source support & Select a focused explanation or demonstration; preserve the distinction between a wrong answer and its possible cause.\\
Retry & Item relationship, identities of returning learners, new and missing respondents & Report a descriptive comparison for the appropriate cohort; avoid attributing a composition change to learning.\\
Plan next lesson & Response evidence, uncertainty, teacher judgment & Decide whether to revisit, gather more evidence or move on; record the reason.\\
\bottomrule
\end{tabular}
\end{table*}

\subsection{Reported classroom setting}
ClassQuest was used with Year 10 students in three approximately 35-minute sessions on 14, 21 and 28 May 2026. Twenty students responded to the usefulness questionnaire from 24 invitees. The account concerns one school setting and computing formative assessment. These May sessions used an earlier or separate prototype, rather than the later implementation snapshot examined below. A contemporaneous October demonstration report describes that later implementation as not yet evaluated with real learners. The classroom account therefore informs the teaching design; it does not validate the audited version or establish that its reporting defects occurred in the May lessons.

The questionnaire concerns understanding concepts, identifying mistakes, and improving answers through feedback and follow-up questions. Selected student feedback refers to Boolean logic, programming loops, explanations, related question practice and partner discussion.

\subsection{Student perceptions of ClassQuest}
Table~\ref{tab:cqratings} gives the aggregate distribution from the classroom usefulness questionnaire. The item asked: ``How useful was ClassQuest in helping you understand computing concepts, identify mistakes and improve your answers through feedback and follow-up questions?'' Its endpoints were 1 (not at all useful) and 10 (extremely useful). The analysis uses the practitioner-supplied aggregate counts; it does not link individual ratings to quiz performance.

\begin{table}[ht]
\centering
\caption{Practitioner-supplied ClassQuest usefulness ratings: 20 respondents from 24 invitees.}
\label{tab:cqratings}
\begin{tabular}{lrrrrrrr}
\toprule
Rating & 1--4 & 5 & 6 & 7 & 8 & 9 & 10\\
\midrule
Count & 0 & 1 & 2 & 4 & 2 & 11 & 0\\
\bottomrule
\end{tabular}
\end{table}

The ratings sum to 160, giving a mean of 8.0. The tenth and eleventh ordered responses are both 9, giving a median of 9. Twenty of 24 invitees responded (83.3\%), and 13 of the 20 respondents rated usefulness at least 8 (65\%). The four non-respondents' views are unknown and are not imputed.

These results indicate favourable perceived usefulness among respondents on the supplied item. The item combines understanding, mistake identification and feedback-supported revision, so it cannot estimate separate perceived effects for those constructs. It also does not establish objective learning, retention, reduced teacher workload, or performance on an independent retry. No participant-level pre/post learning analysis is reported.

\subsection{Qualitative evidence and reporting status}
The practitioner account includes six selected student feedback records, which we paraphrase here. Students described explanations as helping them understand why an answer was incorrect, valued trying a related question after feedback, and found the Boolean-logic quizzes interactive. One response described a benefit from discussing alternative approaches with a partner. Two concerns involved question wording that differed from classroom teaching and insufficient explanation for difficult programming questions, including loops.

These selected comments identify adoption issues without establishing their prevalence. We have not counted theme frequencies or conducted a formal qualitative analysis. In particular, a comment about partner discussion supports a reported experience of that activity, but does not show that a built-in peer interface or the complete structured protocol proposed below was used.

The wording concern suggests a concrete preparation task: review each generated item against the lesson's terminology and learning objective before publication. A technically valid question may still introduce an unfamiliar phrasing that obstructs the intended formative check. The request for more explanation similarly suggests reviewing the rationale, not just the answer key. An educator could pair an explanation with a closely related but distinct item to examine whether the learner can apply the rule. These are implications drawn from the feedback; their effectiveness was not measured in this evaluation.

\subsection{A Boolean-logic teaching episode}
During the 14 May session, students evaluated $A\mathbin{\mathrm{AND}}\mathrm{NOT}\ B$ with $A=\mathrm{True}$ and $B=\mathrm{False}$. Several answered False. The practitioner attributed this difficulty to interpreting how NOT applied to $B$. The teacher used a simple truth table to explain the expression, while ClassQuest provided immediate feedback on incorrect answers. Evaluating the negation gives $\mathrm{NOT}\ B=\mathrm{True}$, so the full expression is True.

Students then attempted a fresh item, $X\mathbin{\mathrm{OR}}\mathrm{NOT}\ Y$, with $X=\mathrm{False}$ and $Y=\mathrm{False}$. This also evaluates to True. The practitioner reported that most students subsequently applied the operators correctly. No response count or matched individual result accompanies that observation, so we report it as a qualitative classroom account rather than a measured gain. The outer operator changes from AND to OR, both correct answers are True, and feedback and teacher explanation occur together; this episode cannot establish equivalent item difficulty, retained understanding or an effect attributable to ClassQuest alone.

The episode illustrates an adoptable teaching response: make the negated subexpression explicit before evaluating the complete expression. A future follow-up could ask learners to explain that intermediate step and include items with different final truth values. These are recommendations arising from the episode, not additional activities claimed to have occurred.

\subsection{Ethics and evidence boundaries}

Ethics approval was obtained for the classroom research. Written parental or guardian consent and student assent were obtained before research data collection. Participation was voluntary, and students could complete normal classroom activities without contributing data. Evaluation records used participant identifiers, with personal information and consent documents stored separately under restricted access.

The questionnaire describes student perceptions rather than a measured learning gain. The technical findings below provide a separate source of evidence, using explicitly constructed data to examine reporting behaviour.

\section{Technical audit method}
We froze the reporting and question-validation modules and recorded their SHA-256 hashes. The original 5,456-state audit ran with Python 3.11.11 on a Linux research VM without model inference. The later opportunity-to-respond case and companion checks ran locally on a CPU, also without model inference. These tests concern the identified snapshots, rather than all versions of the application. Their inputs are artificial; no human participants supplied responses to the software audits.

For one item and a roster of 30, we enumerated every non-negative count tuple $(c,w,p,m)$ satisfying
\[
 c+w+p+m=30,
\]
where $c$ is correct, $w$ incorrect, $p$ pending review and $m$ missing. There are $\binom{33}{3}=5,456$ such configurations. An independent arithmetic oracle checks the report's rounded accuracy $100c/(c+w)$ when at least one answer is scored, its undefined value when none is scored, and the pending and missing counts.

The existing topic recommendation uses provisional 55\% and 80\% accuracy bands and an early-signal rule below three scored answers. These are implementation heuristics, not validated pedagogical thresholds. For configurations with at least three scored answers, we also compare categories under the two extreme completions of pending and missing outcomes: all incorrect or all correct. This explores sensitivity to unobserved outcomes, not a probabilistic model of how learners would respond.

Additional diagnostic cases examine a changing retry cohort and several answers supplied by one learner. Source-validation cases vary a quote, locator, prompt or answer key while holding other properties fixed. A deliberately wrong short explanation includes rubric words to test whether concept overlap is being treated as final correctness. The cases identify specific mechanisms; their frequency in schools is not measured.

\section{Audit findings and their teaching implications}
\subsection{An unoffered item is not a missing response}
An October demonstration report identified an inactive-retry reporting problem, which we reproduced in the frozen reporting module. In a constructed three-person room, all three learners answer the first item and its selected Boolean-logic retry. A second retry belongs to an unselected Loops topic. When the full question list and the room's selected retry topic are supplied, the report labels all three learners missing for the Loops item. The reporting function does not use that selection to establish who was offered each question.

This is a different problem from an absent answer to an assigned item. Reteaching loops or prompting learners to complete the task would be poorly grounded when the task was never offered. Filtering the question list upstream can avoid this specific case; recording item-level opportunity is necessary when different learners receive different tasks. The audit establishes the behaviour of the frozen function under these inputs, not its prevalence in classrooms.

\subsection{Correct arithmetic does not settle coverage}
All 5,456 single-item configurations matched the arithmetic oracle. This provides evidence that the audited counting logic handles the enumerated states as intended. It does not establish that a recommendation based on those counts is justified.

Of the 5,278 configurations with at least three scored answers, 4,141 crossed a recommendation category between the two extreme completions of unknown outcomes. These counts come from an enumerated artificial state space and should not be reported as a classroom error rate. The practical issue is simpler: some recommendations depend on outcomes that the teacher has not yet observed or scored.

Consider an illustrative item with 12 correct, three incorrect, five pending and ten missing responses in a 30-person roster. Observed accuracy is 80\%, but only half the roster has a scored answer. Pending work already exists and needs judgment; missing work does not. These states call for potentially different actions. Reviewing five explanations and checking whether ten students could participate are more informative than silently treating all 15 unknown outcomes as wrong or ignoring them.

This illustrative item is not a classroom observation. It makes the reporting requirement concrete: display the numerator, scored denominator, roster size, pending count and missing count together. A teacher can then judge whether to gather more evidence before deciding on another explanation.

\subsection{A retry increase caused by participation}
In a constructed 20-person first round, four answers are correct. Only these four learners submit a retry and all remain correct. The current report compares 20\% with 100\% and displays an increase of 80 percentage points. However, the four returning learners changed by zero percentage points; the other 16 supplied no retry evidence.

The discrepancy is not an arithmetic bug. It is a comparison of different groups. A teacher seeing only the increase could infer that an explanation succeeded when the report provides no retry evidence from the initially incorrect group. The reporting remedy is to expose the matched cohort and the learners observed in only one round.

Matching respondents is necessary for this descriptive question, but it is not sufficient for a learning claim. The retry may be easier, repeat the same answer, or follow an explanation that temporarily cues the response. The guide therefore treats a matched result as descriptive and asks whether the item demands are comparable before making a teaching interpretation.

\subsection{Answers are not distinct learners}
A second constructed case gives one learner three correct answers in a 30-person roster. The topic report counts three scored answers and recommends moving on. Its early-signal rule is satisfied by answer count, although the evidence comes from only one person.

For a topic containing multiple items, a report should show both scored-answer count and distinct respondents. These quantities answer different questions. A teacher may value a detailed response pattern from one learner while still lacking evidence about the class. Adding more questions to the same person's record should not silently create the appearance of broader participation.

\subsection{Source membership and answer validity}
The structural validator accepted a genuine source quote paired with a wrong answer key, and accepted a genuine quote paired with an unrelated question. It rejected a fabricated quote and an incorrect locator. This behaviour shows what substring validation can establish: that the quoted text occurs at the stated location. It does not establish entailment between the passage, prompt and answer.

A deliberately incorrect short answer containing all rubric terms received a maximal overlap hint, but correctly remained pending review. This is evidence that the review boundary is preserved in that case. It is not evidence of accurate automated semantic grading. Teachers should inspect the explanation itself, including negation, causal direction and the use of examples.

\begin{table}[ht]
\caption{Completed diagnostic findings; all audit responses are constructed.}
\begin{tabular}{@{}p{0.35\columnwidth}p{0.59\columnwidth}@{}}
\toprule
Check & Result\\
\midrule
Unoffered retry & Three missing labels although nobody was offered the item\\
Single-item accounting & 0 mismatches in 5,456 configurations\\
Selective retry & Displayed +80 points; matched change 0 points, four learners\\
Repeated items & Three correct answers from one learner trigger progression suggestion\\
Real quote, wrong key & Accepted by structural validation\\
False quote or locator & Rejected\\
Wrong explanation with rubric terms & Maximal overlap hint; remains pending human review\\
\bottomrule
\end{tabular}
\end{table}

\section{Reporting companion and decision guide}
\subsection{Implemented reporting additions}
A separate Python companion computes roster size, distinct respondents, scored coverage, pending and missing counts, and observed item accuracy. For one item with roster size $N$, it also reports
\[
 \left[\frac{c}{N},\frac{c+p+m}{N}\right].
\]
These are logical completion bounds obtained by making every unknown outcome incorrect or correct. They are not confidence intervals, estimates of mastery, or a recommendation to impute unknown work. For the illustrative 30-person item above, the bounds are 40\% to 90\%, compared with 80\% accuracy among the 15 scored answers.

For a declared first/retry item pair, the companion matches respondent identifiers and reports matched count, first-only count, retry-only count and descriptive matched change. It rejects duplicate participant records within the item input. In the selective-retry case it reports four matched learners, 16 first-only learners and zero matched change. The separate companion is implemented and exercised on the constructed examples; these additions are not yet integrated into the classroom interface or evaluated with teachers.

\subsection{Recording who was offered the item}
The companion now accepts an explicit set of learners offered a particular item, as a subset of the class roster. It reports \emph{not offered} separately from \emph{offered but unanswered} and \emph{answered but awaiting review}. Accuracy and completion bounds concern the offered subset. In the constructed inactive-retry case, it returns three not offered, zero missing and undefined accuracy. It refuses an unknown offer set rather than infer exposure from the responders, and rejects responses from learners outside that set. An offer record means an item was made available; it does not prove that a learner read it or could access it successfully.

We checked all 19,531 assignments for rosters of zero to six learners, using five states per learner: not offered, missing, pending, correct and incorrect. All matched an independent state-count oracle. Adding a learner who was not offered the item left the item accuracy, coverage and completion bounds unchanged in every case. Seven invalid-input cases were rejected. These are software properties of the companion, not student outcomes. The practical requirement is an auditable assignment record; where it is unavailable, the teacher should treat the reason for absence as unknown.

\subsection{An executable check before interpreting a band}
The companion also computes which reporting bands remain possible when pending and missing responses are resolved. It enumerates integer correct totals from $c$ through $c+p+m$, applies the frozen report's nearest-integer, ties-to-even percentage rounding, and retains every attainable 55/80 band. It flags a completion-sensitive result when more than one band is possible. The early-signal flag below three scored responses remains separate. The 55/80 cutoffs are reproduced implementation rules, not recommendations about mastery.

For the 12-correct example, reteach, recap and advance are all attainable; a displayed advance label therefore depends on unresolved outcomes. The teacher sees five pending explanations and ten missing responses alongside that flag. Conversely, a stable band only means that completing this binary item cannot change that particular heuristic. It does not establish a valid question, adequate explanation, representative participation or a uniquely appropriate teaching action. The empty-roster case produces no band.

We tested all 46,375 non-negative count configurations across roster sizes 1--30 against an independent integer-inequality oracle that enumerates every possible completed total. There were zero mismatches. Resolving one pending or missing outcome either way cannot introduce a previously impossible band; all 163,680 applicable one-response transitions satisfied this property. These checks cover the new companion, separately from the 5,456-state audit of the frozen application. The earlier 30-person sensitivity count is unchanged by the rounding convention because its attainable full-roster percentages do not lie at a differing boundary.

This test also matters for small classes: one unresolved learner in a one-person roster permits reteach or advance, but no attainable total falls in recap. The companion reports only attainable bands, while retaining the early-signal flag. The mechanism is a reproducible guard against interpreting incomplete item evidence as a settled recommendation; its effect on teacher decisions has not been evaluated.

\subsection{A practical interpretation sequence}
We recommend beginning with participation rather than an action label. First, establish the relevant roster, which learners were offered each item, and whether they could respond. Second, distinguish work needing human review from missing work. Third, inspect the response pattern and any supporting explanation. Fourth, if comparing rounds, identify the matched group and the relationship between the items. Only then select a teaching action.

Several actions may be defensible for the same numerical result. A teacher might review pending explanations, ask an additional individual question, revisit vocabulary, demonstrate a concept, or move to a new task while supporting a smaller group. The report should make the basis for that judgment visible rather than imply that a percentage uniquely determines the pedagogy.

A brief teacher note can record the concept, the evidence considered, the uncertainty, the chosen action and the follow-up evidence to collect. This is a proposed practice for using the companion, not a measured intervention. Its value should be evaluated against actual preparation and classroom workload, rather than assumed from the number of fields displayed.

\subsection{Choosing an action from the unresolved evidence}
The checks have a practical order. Establish that an item was offered before interpreting absence; establish that a response is scored before using accuracy; establish who answered both rounds before interpreting change. Then inspect whether the item and explanation actually address the learning objective. A failure at one step identifies evidence to obtain, rather than automatically prescribing a teaching action.

For example, a teacher facing a low response count should check assignment and access before assuming disengagement. A high accuracy with pending explanations calls for reviewing those explanations before concluding that the concept is secure. A large retry increase with a changing cohort calls for a matched description and fresh evidence from the initially incorrect learners. These different responses follow from the audit cases even when the interface supplies the same broad suggestion to gather more answers. They can be applied with a class register and spreadsheet without installing ClassQuest.

\subsection{A decision record another teacher can use}
The protocol can be adopted with a spreadsheet or paper record as well as the companion. For each item or explicitly defined item set, record the intended concept, the reviewed source and key, the number offered the item, distinct respondents, correct/incorrect/pending/missing counts, the observed difficulty, the selected response and the evidence to collect next. A retry entry additionally records who answered both items and why the items address the same concept. These fields specify an interpretation task, not a validated scale or an automated pedagogical prescription.

For the constructed 30-person example, a defensible note would be: ``15 scored responses; 12 correct; five explanations pending; ten learners missing. Review the explanations and establish whether missing learners could participate before deciding whether the class is ready to progress.'' For the reported Boolean lesson, the corresponding next-evidence question is whether a learner can explain the negated subexpression on a fresh item. The first note is a worked audit example; the second is an implication of the classroom episode. Neither invents a recorded teacher decision.

Before adopting the workflow, an educator needs a short source aligned with a computing objective, reviewed items and rationales, a way to record individual participation, and time to examine pending explanations. Model generation is optional. If source alignment is uncertain, revise or discard the item; if participation is incomplete, obtain the missing evidence; if the explanation is unclear, inspect reasoning before using a correct-answer percentage. These actions give the audit practical value even when its companion is not installed.

\section{Extending the activity to peer learning}
The practitioner's feedback includes a request to extend peer learning, and a selected student response describes discussion with a partner. This establishes a reported classroom practice, not implementation of a structured digital peer workflow. We therefore specify the fuller source-checked explanation and revision design below as an extension.

In a source-checked peer explanation activity, each learner first commits an answer and confidence privately. Partners then explain their reasoning using a located source passage. Each asks the other to address an alternative interpretation or counterexample. Learners record a revision, if warranted, before attempting a fresh question individually. A teacher remains responsible for reviewing ambiguous source interpretations and closing the discussion.

A proposed concept-repair record would retain the original answer, source used, peer challenge, revised explanation and independent retry. Its purpose is to distinguish answer switching from a change in reasoning. Correct-to-incorrect revisions and unsupported agreement should remain visible. The record should not label every revision as improvement or infer understanding from agreement.

For a Boolean-logic lesson, a pair might explain why a conjunction is false when one operand is false, then test that explanation against a different truth assignment. This is a proposed instructional example, not a reported classroom episode. A student-authoring variant could require an answer rationale and source alongside a misconception-based distractor, but it builds on established student-question authoring~\cite{peerwise} rather than introducing that pedagogy.

\section{Limitations and evaluation priorities}
The audit concerns selected mechanisms in one frozen prototype. The frozen-module census covers one item at roster size 30; the completion check covers one item at roster sizes 1--30. Neither covers every multi-item or deployment situation. Constructed cases cannot estimate classroom prevalence. The original and companion reports have not been compared in a teacher decision study, and the proposed peer workflow has not been shown to improve explanations or transfer.

The classroom evidence comes from one setting and consists of aggregate perceptions, selected feedback and a practitioner-reported teaching episode; it does not establish learning gains or reduced teacher workload. The later software audit concerns a different prototype snapshot and cannot establish which mechanisms occurred during those lessons. The opportunity check assumes reliable assignment records; its exhaustive enumeration covers at most six learners and does not establish accessibility or successful delivery.

A next study could ask teachers to interpret counterbalanced, educator-reviewed computing scenarios using the existing and companion reports. The principal outcome should concern the defensibility of the chosen action, scored with a rubric that allows more than one sound response. Decision time and perceived usefulness would provide complementary evidence. This is a proposed evaluation, not an experiment reported here.

A later classroom investigation could examine explanations and fresh individual tasks after peer discussion, using comparable items, matched learners and a documented teacher intervention.

Generative AI (OpenAI ChatGPT/Codex) assisted literature discovery, code development, source inspection and manuscript drafting and revision. The deterministic software audit used constructed inputs, not human-participant responses. Classroom results are reported separately from the practitioner account.

\section{Conclusion}
ClassQuest connects source-based question review to formative teaching decisions. Students reported favourable perceptions alongside concerns about wording and explanation depth. The Boolean-logic episode illustrates feedback, teacher explanation and a fresh question in classroom practice, without isolating a learning effect. The separate audit shows why a correct percentage alone is insufficient: unoffered tasks, unresolved responses, changed retry participation and repeated items can alter the interpretation. The reporting companion and adoption guide expose those boundaries, and the tested completion check identifies recommendations sensitive to unresolved responses. Their educational effects remain to be evaluated.
\bibliographystyle{ACM-Reference-Format}
\bibliography{references}
\end{document}